# SPEED OPTIMIZATION IN UNPLANNED TRAFFIC USING BIO-INSPIRED COMPUTING AND POPULATION KNOWLEDGE BASE

Prasun Ghosal[1], Arijit Chakraborty[2], Sabyasachee Banerjee[2], Satabdi Barman[2]

[1] Department of Information Technology,
Bengal Engineering and Science University, Shibpur, Howrah, W.B., India,
(e-mail:p_ghosal@it.becs.ac.in)
[2] Department of Computer Science & Engineering,
Heritage Institute of Technology, Kolkata, W.B., India,
(e-mail: {arijitchakraborty.besu, sabyasachee.banerjee,
satabdi.barman}@gmail.com)

## Abstract

*Bio-Inspired Algorithms on Road Traffic Congestion and safety is a very promising research problem. Searching for an efficient optimization method to increase the degree of speed optimization and thereby increasing the traffic Flow in an unplanned zone is a widely concerning issue. However, there has been a limited research effort on the optimization of the lane usage with speed optimization.*

*The main objective of this article is to find avenues or techniques in a novel way to solve the problem optimally using the knowledge from analysis of speeds of vehicles, which, in turn will act as a guide for design of lanes optimally to provide better optimized traffic. The accident factors adjust the base model estimates for individual geometric design element dimensions and for traffic control features. The application of these algorithms in partially modified form in accordance of this novel Speed Optimization Technique in an Unplanned Traffic analysis technique is applied to the proposed design and speed optimization plan. The experimental results based on real life data are quite encouraging.*

## 1. INTRODUCTION

Bio-Inspired algorithms present a new optimal lane analysis as a guide for designing of non accidental lane to serve better utilization of lane. The accident factors adjust the base model estimates for individual geometric design element dimensions and for traffic control features.

Bio inspired analysis technique is applied to the proposed design and speed optimization plan. Design of Non Accidental Lane can robustly manage and operations on lane for avoiding accident. Therefore how to increase the Speed optimization with non accidental zone of the Lane is widely concerting issue. There has been a limited research effort on the optimization of the DNAL systems.

                                        79

In spite of limitation of this algorithm it can be clearly shown that we can optimize speed in lieu of number of lane transition. Before we go to further details let me describe the background behind this idea in the following paragraphs.

Bio-Inspired Algorithms are inspired by a variety of biological and natural processes that had been observed over years. The popularity of the Bio-Inspired Algorithms is primarily caused by the ability of biological and natural systems to effectively adjust to frequently changeable environment.

For e.g.: Evolutionary computation, neural networks, ant colony optimization, particle swarm optimization, artificial immune systems, and bacteria foraging algorithm are the algorithms and concepts that were motivated by nature.

Swarm behavior is one of the main characteristics of different colonies of social insects (bees, wasps, ants, termites). This type of behavior is first and foremost characterized by autonomy, distributed functioning and self-organizing. Swarm Intelligence [Beni and Wang 1989] is the area of Artificial Intelligence that is based on study of actions of individuals in various decentralized systems. When creating Swarm Intelligence models and techniques, researchers apply some principles of the natural swarm intelligence.

## 1.1 Historical Background: Ant Colony System

Basic flow of an ant colony based system may be represented with the following figure.

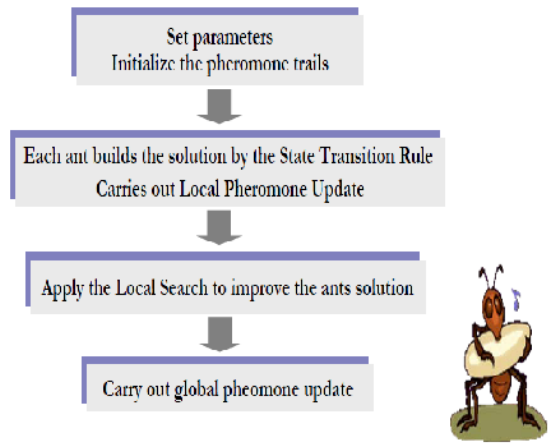

# 2 BACKGROUND AND MOTIVATION

## 2.1 Related Works

Jake Kononov, Barbara Bailey, and Bryan K. Allery, first explores the relationship between safety and congestion and then examines the relationship between safety and the number of lanes on urban freeways. The relationship between safety and congestion on urban freeways was explored with the use of safety performance functions [SPF] calibrated for multilane freeways in Colorado, California, and Texas.

The Focus of most SPF modeling efforts to date has been on the statistical technique and the underlying probability distributions. The modeling process was informed by the consideration of the traffic operations parameters described by the Highway Capacity Manual. [1]

H Ludvigsen, Danish Road Directorate, DK; J Mertner, COWI A/S, DK , 2006, explored Differentiated speed limits allowing higher speed at certain road sections whilst maintaining the safety standards are presently being applied in Denmark. The typical odds that higher speed limits will increase the number of accidents must thus be beaten by the project. The Danish Road Directorate has been asked by the Ministry of Energy and Transport based on a request from parliamentarians to suggest an approach to assess the potential for introduction of differentiated speed limits on the Danish state road network. A pilot project was carried in late 2006 and the entire state network will be assessed during the first half of 2007 - first of all to identify where speed limits may be raised. The paper will present the methodology and findings of a project carried out by the Danish Road  Directorate and COWI aimed at identifying potential sections where the speed limit could be increased from 80 km/h to 90 km/h without jeopardizing road safety and where only minor and cheaper measures are necessary. Thus it will be described how to systematically assess the road network when the speed limit is to be increased... [2]

C.J. Messer and D.B. Fambro, 1977, presented a new critical lane analysis as a guide for designing signalized intersections to serve rush-hour traffic demands.

Physical design and signalization alternatives are identified, and methods for evaluation are provided. The procedures used to convert traffic volume data for the design year into equivalent turning movement volumes are described, and all volumes are then converted into equivalent through-automobile volumes.

The critical lane analysis technique is applied to the proposed design and signalization plan. The resulting sum of critical lane volumes is then checked against established maximum values for each level of service (A, B, C, D, E) to determine the acceptability of the design. [3].

In the Operation and Safety of Right-Turn Lane Design's objectives of this research by the Texas Department of Transportation were to determine the variables that affect the speeds of free-flow turning vehicles in an exclusive right-turn lane and explore the safety experience of different right-turn lane designs. The evaluations found that the variables affecting the turning speed at an exclusive right-turn lane include type of channelization present (either lane line or raised island), lane length, and corner radius. Variables that affect the turning speed at an exclusive right-turn lane with island design include: (a) radius, lane length, and is land size at the beginning of the turn and (b) corner radius, lane length, and turning-roadway width near the middle of the turn. Researchers for a Georgia study concluded that treatments that had the highest number of crashes were right-turn lanes with raised islands. This type of intersection had the second highest number of crashes of the treatments evaluated in Texas. In both studies, the "shared through with right lane combination" had the lowest number of crashes. These findings need to be verified through use of a larger, more comprehensive study that includes right-turning volume. [4]

## 2.2 Drawbacks of existing solutions

Many traditional speed-optimizing algorithms for lanes were proposed earlier to optimize deterministic problems. But these algorithms didn't show their ability to use their previous knowledge to tackle the inherent randomness in the traffic systems. Therefore, to handle with such random realistic situation and generate some efficient solution, good computational models of the same problem as well as good heuristics are required.

This article is divided into two major sections: -

In first part, simulation algorithm will provide us with no. Of lanes required moving the traffic at optimal speed in each proposed lane.

Second part, deals with knowledge obtained from the first part to make the lane transitions less in number making it nearer towards the desired goal.

# 3 PROBLEM FORMULATION AND PROPOSED ALGORITHMS

## 3.1 Problem Description

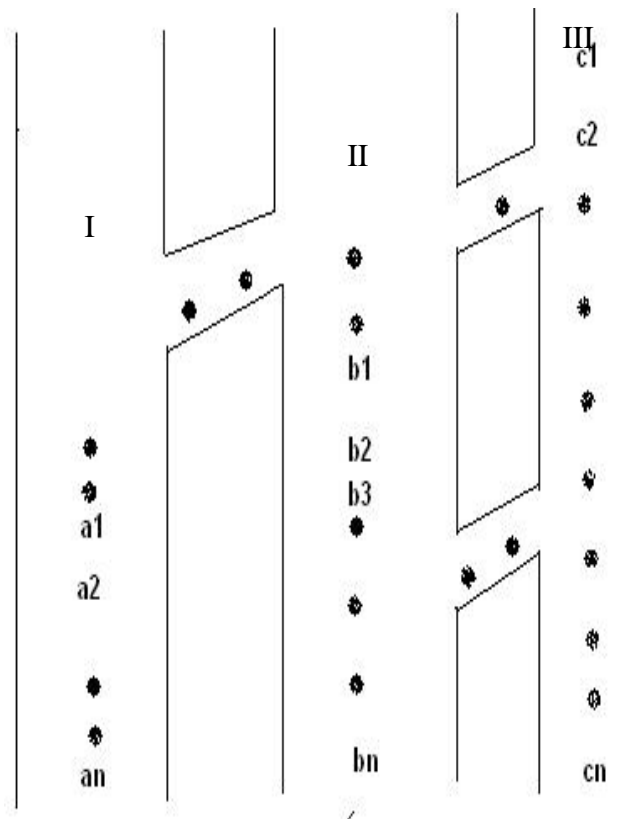

Figure 1: Mechanism of Lane Transition

**Description of Figure 1:**

Figure 1, Three vertical lanes that are unidirectional, and A = {a1, a2… an}, B = {b1, b2 ….bn}, C = {c1, c2,….,cn}, three lanes. I, II, III are the transition points through which vehicles can overtake its preceding vehicle with lesser speed and then immediately moves to its original lane. i.e. I from lane A to B or B to A and II, III are from B to C or C to B. Here we assume that each and every lane's car speed is greater than 0 kmph. If speed of any car is less than or equal to 0 kmph then we assume that there may be problem.

## 3.2 Problem Formulation

Random movement of vehicle in rush hour traffic are required to be frame up in optimal no. lanes with respect to number of transitions between lanes so that each lane have optimal speed.

Bio inspired algorithms like swarm intelligence (.i.e. Ant Colony Optimization) technique used here with speed of the 'vehicle' acting as a pheromone to solve the problem in hand.

To maintain the optimality of a solution in a heuristic search using population information as a knowledge base is used in the proposed algorithms.

## Assumption

During the execution of the algorithm it is assumed that there will be no change in the current speed of the vehicle whenever the current speed of the vehicle is entered once

## Description of the Proposed Algorithm

The primary sections of the proposed algorithms and their major functionalities are described below.

Step 1:
Taking input from sensors, like the current speed of the vehicle, arrival time etc., and, counting the number Vehicles the user has entered.
Step 2:
Categorizing the vehicles depending on their Current speed.
Step 3:
Checking total how many numbers of lanes will be required for our sample data in an unplanned zone, and, which vehicle is moving in which lane.
Step 4:
Checking total number of transitions i.e. at which point of the lane and from which lane to where the transition will occur. Now these are the outputs of our simulation.

**Symbolic Interpretation used in the Algorithms:**

| Symbols used | Meaning |
|---|---|
| $V_i$ | Velocity of vehicle i |
| $V_j$ | Velocity of vehicle j |
| $L_i$ | Lane of the vehicle i |
| $L_j$ | Lane of the vehicle j |
| $L_1$ | Lane of the 1st vehicle. |
| type(i) | Category of Vehicle i |
| t | Arrival time difference between a high and low speed vehicles |
| t1 | Time interval to overtake a vehicles at lower speed |
| d | Distance covered by low speed Vehicle |
| d1 | Distance covered by high speeding Vehicle |
| $B_n$ | Buffer of Lane n |
| Count | Total no. Vehicle in unplanned traffic |
| Count1 | Total no. Lanes for optimal speed |
| Count2 | Total number Of transition |

### 3.2.1 Algorithm Part I:

Input: Details of vehicles, Current speed of the vehicle, arrival time.
Output: Category of the vehicle, Number of lanes will be required, Number of transitions.
Step 1.1: Set count = 1; /*Used to count the number of vehicles. */
Step 1.2: get_ input (); /*Enter Details of vehicles, current speed, arrival time and store it inta record. */
Step 1.3: Continue Step 1.1 until sensor stops to give feedback and
        Update count = count + 1 for each feedback;
Step 2: For 1 <= i <=count for each vehicle
If 0 < Vi <11    then categorize Vi   as type A
If 10< Vi < 31  then   categorize $V_i$   as type B
If 30< Vi < 46  then   categorize $V_i$   as type C
If 45< Vi < 51  then   categorize Vi   as type D
If 50< Vi < 101 then   categorize $V_i$   as type E
Step 3: Set counter: count1: = 1;
Set $L_1$= 1;
        For 2 <= i <= count for each Vehicle
            For 1 <= j <= count1
            Compare the {type(i) , type(j )}  present in the lane
            If different update count1 = count1 + 1 and
            $L_i$= count1;
            Else

$L_i = j$;
                End of loop;
        End of loop;
Step 4: Set counter: count2 = count1;
        For $1 <= i <= $ count -1 for each Vehicle
`       For $2 <= j <= $count for each Vehicle
                If type($V_i$)= type( j) and Vi < Vj and arrivaltime($V_i$) <= arrivaltime($V_j$)
                Set t =arrivaltime ($V_j$) – arrivaltime (Vi);
                Set t1 = 0;
                Begin loop
                        Set t1 = t1 + 1;
                        Set d = Vi x (t + t1);
                        Set d1 = Vj x t1;
                        If d1 <= d Set count2 = count2 + 1;
                        If $L_j$ = 1 then transition will be to 2 - lane;
                        If $L_j$ = count1 then transition is count1 - lane;
                        Else

                                Transition is Lj - 1 or Lj + 1;
                End loop;
        End loop;
      End loop;
Step 5: Return Number of lanes required = count1;
        Number of transitions required = count2;
Step 6: End

### 3.2.1.1 Analysis of the Proposed Algorithm (Part I)

- The above algorithm is implemented on an open unplanned Area.
- The objective will follow linear queue as long as speed/value/cost of proceeding is greater than the immediate next.
- Transition/Cross over are used and they again follow appropriate data structure in order to maintain the preceding step rule.
- Here we assume the lanes are narrow enough to limit the bi-directional approach.
- Here we maintain optimize speed for each lane.
- Here we also maintain the transition points if speed/value/cost of a vehicle is found unable to maintain the normal movement and transition in all the calculated lanes.
- Transition points are recorded with their position and number and it follows appropriate data structure in order to maintain the record.

### 3.3. Proposed Algorithm (Part II)

**Description of the proposed algorithm.** The primary sections of the proposed algorithm and their major functionalities are described below.

- Step 1. Take relevant information from sensors, i.e. the current speed of the vehicle, arrival time etc. and count the number of vehicles the sensor has entered along with that consider number of lanes that are present in the traffic.
- Step 2. Assign lanes to different vehicles having different current speeds at any time instant t in order to categorize them.
- Step 3. Determine whether the current speed of the vehicle is equal to the speeds present in speed buffers of lanes or not.
- Step 4. This step finds the lane, where, the difference between the vehicle's current speed and lane's speed buffer's average speed is minimum and takes the vehicle to the lane, categorizes it same as the lane's other vehicles, increases the population of the lane, and stores the vehicle's current speed in the speed buffer of the lane.
- Step 5. This step is used for checking total numbers of transitions, i.e. at which point of the lane and from which lane to where the transition will occur, thereby calculating the average speed of the lanes.

### 3.3.1 Pseudo code of the proposed algorithm. (PartII)

INPUT: Vehicle's name, current speed, arrival time.
OUTPUT: Vehicle's Type, Number of transitions.
Step 1.1: Set count=1; /*used to count the number of vehicles*/
Step 1.2: get_input ()/*Enter the inputs when speed of the vehicle is non-zero. */
Step 1.3: Continue Step 1.1 until sensor stops to give feedback.
Step 2: Set type(1 )='A', Enter $V_1$ into $1^{st}$ lane's speed buffer, Set 1st lane's population (count_l) as '1', Set n=2.
For 2 i count
Set a buffer buf=0
Loop1 until lane='0'
Loop2 for 1 j<I for each vehicle
If Vi = Vj
Set buf =1, type (i) =type (j)
Goto Step 3 and send Vi to Step 3 as 'speed1'.
Step 2.1
If buf=1 then end Loop1
If buf=0
Enter Bn=Vi, Set count_l=1, Set type (i )=A++;
End Loop1          /*Bn=n lane speed buffer*/
If lane=0
Then end Loop1.
If lane=0
Then end Loop.
Store buf2=i+1
Step 3: For 1 i lane_l for each lane

If Bi's 1$^{st}$ speed=speed1
Update count_li++;
Set Bi, count_1=speed1
goto step 2.1
Step 4: for buf2  i  count
Set c=1, switch=0.
Set min=|Vi, Lc|, /*Lc=c lane's average speed*/
type (i) =1st lane's vehicle type
For 1  j  lane_1
Set d=|Vi, Lj|
If d=0
Set type (i) = type (L$_j$)
Update (j) lane's count_l= (j) lane's count_l+1
Set switch=1
End Loop
If d<min
Then min=d
Set type (i) = type (L$_{j)}$
Update (j)$^{th}$ lane's count_l= (j) lane's conut_l+1
Set (j)$^{th}$ lane's speed buffer [count_l] = (i) vehicle's speed (Vi)
If switch=0
Update L1, count_1 ++;
Step 5: Set count2 as count2 =1
For 1  i  count-1
For 2  j  count
If type (i) = type (j) and Vi<Vj and (i) vehicle's arrival time   (j) vehicle's arrival time
Set t= (j) vehicle's arrival time - (i) vehicle's arrival time
Set t1=0
Begin loop
Set t1=t1+1
Set d=Vi*(t+t1)
Set d1=Vj*t1
If d1  d set count2 = count2+1
If Lj =1 then transition will be to 2-lane
If Lj =count1 then transition will be to count1-lane
Else transition will be to L$_j$ -1 or L$_j$ +1
End loop
End loop
End loop
For 1  m  lane_1
Calculate each lane's average speed from its speed buffer.
Step 6: Return Number of transitions required= count2
Step 7: End.

### 3.3.1.1 Analysis of the proposed algorithm.

The salient points and features of the proposed algorithm may be analyzed as follows.

- The above algorithm is implemented on an open lane area.
- The objective will follow linear queue as long as speed/value/cost of proceeding to greater than the immediate next.
- Transition/Cross over are used and they again follow appropriate data structure in order to maintain the preceding step rule.
- Here we assume the lanes are narrow enough to limit the bidirectional approach.
- Here we also maintain the transition points if speed/value/cost of a vehicle is found unable to maintain the normal movement and transition in all the calculated lanes.
- Transition points are recorded with their position and number and it follows appropriate data structure in order to maintain the record.

# 4 EXPERIMENTAL RESULTS AND OBSERVATIONS

The optimization of the speed in rush hour traffic with the swarm intelligence approach in an open lane area used the population information as a knowledge base. Primary objective of this approach is to improve the traffic movement in rush hours and to optimize the speed of the vehicles using the concept of transition points between adjacent Lanes.

The above proposed algorithms has been implemented using programming language ANSI C in an open platform, on a Intel Pentium IV processor with 1 GB physical memory.

## 4.1 Simulated Graphical Analysis of the proposed Algorithms

By implementing the above proposed algorithm and doing the simulation we were able to generate the following graphical results shown in figures 2 and 3 as follows

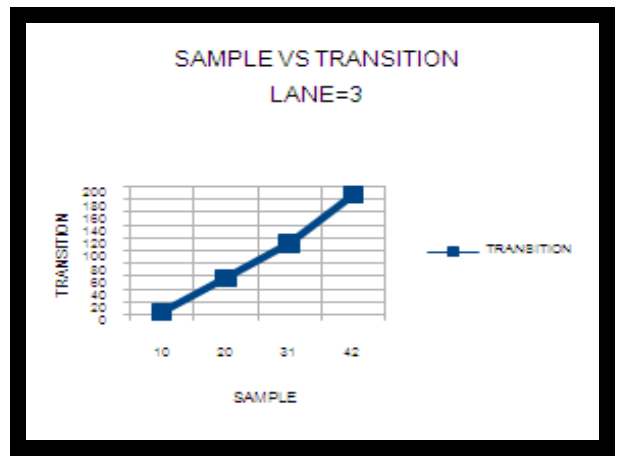

Figure 2: Transitions increases linearly with sample size

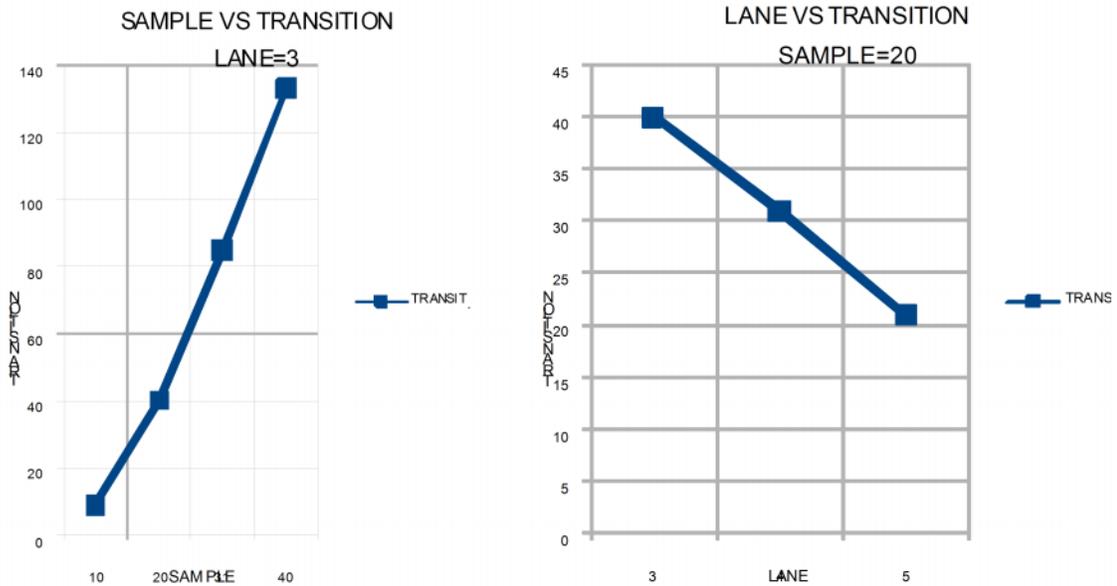

Figure: Graphical Analysis between swarm intelligence and population based optimization

# 5 ANALYSIS WITH ROAD TRAFFIC DATA

Specimen Data Tables: Transport

| Total Registered Motor Vehicles in Metropolitan Cities (As on 31.03.1996) | | | | | | | | | | | | |
|---|---|---|---|---|---|---|---|---|---|---|---|---|
| | | | | | | | | | | | | (In Number) |
| Metropolitan Cities | All Vehicles | Two Wheelers | Three Wheelers | | Cars | Jeeps | Taxis | Buses | Trucks | Tractors | Traillers | Others |
| | | | Passenger | Goods | | | | | | | | |
| Ahmedabad | 571643 | 441919 | 34228 | 6857 | 56557 | 4632 | 3653 | 14447 | 8319 | 97 | 154 | 780 |
| Bangalore | 900541 | 669937 | 39292 | 7913 | 121408 | 5996 | 3331 | 11655 | 21826 | 3872 | 3661 | 11650 |
| Bhopal | 222619 | 158623 | 7063 | 1229 | 13715 | 15614 | 1110 | 3565 | 7906 | 8558 | 3418 | 1818 |
| Calcutta | - | - | - | - | - | - | - | - | - | | | |
| Chennai | 811916 | 593750 | 25320 | 2611 | 145222 | 5944 | 433 | 9912 | 21126 | 382 | 374 | 6842 |
| Cochin | 196559 | 113749 | 14287 | 2757 | 28656 | 3236 | 5692 | 5605 | 19775 | 443 | 1346 | 1013 |
| Coimbatore | 240610 | 177061 | 7218 | 1007 | 31307 | 2555 | 60 | 3273 | 10128 | 3551 | 1830 | 2620 |
| Delhi | 2629645 | 1741776 | 79011 | - | 633852 | A | 13663 | 27871 | 133472 | - | - | - |
| Hyderabad | - | - | - | - | - | - | - | - | - | | - | - |
| Indore | 324740 | 241780 | 8772 | 1604 | 29481 | 3824 | 1136 | 5128 | 21231 | 6027 | 4491 | 1266 |
| Jaipur | 405499 | 300294 | 6040 | - | 34614 | 12070 | 3331 | 11673 | 20379 | 14116 | 2669 | 313 |
| Kanpur | 246801 | 194547 | 4644 | 1057 | 14827 | 3354 | 579 | 1635 | 8730 | 15527 | 1014 | 887 |
| Lucknow | 303356 | 242946 | 7346 | 2215 | 23917 | 7754 | 2462 | 1750 | 4553 | 7453 | 822 | 2138 |
| Ludhiana | - | - | - | - | - | - | - | - | - | - | - | - |
| Madurai | 116765 | 89169 | 2896 | 1477 | 7329 | 369 | - | 3048 | 6339 | 2238 | 786 | 3114 |
| Mumbai | 723632 | 302513 | 59222 | 25327 | 234853 | 17845 | 44842 | 12277 | 19357 | 1120 | 1143 | 5133 |
| Nagpur | 213404 | 173154 | 6918 | 4570 | 13187 | 4156 | 1225 | 1728 | 7263 | 292 | 494 | 417 |
| Patna | 219533 | 150979 | 10757 | 243 | 19559 | 9817 | 2245 | 3202 | 12193 | 5255 | 4586 | 697 |
| Pune | 411880 | 311748 | 26222 | 5145 | 34496 | 6232 | 2601 | 5814 | 17713 | 572 | 595 | 742 |
| Surat | 330961 | 282656 | 13738 | 3060 | 25658 | 1991 | 669 | 559 | 2068 | 110 | 172 | 280 |
| Vadodara | 275473 | 208792 | 16129 | 6788 | 28211 | 4578 | 2669 | 1759 | 3692 | 968 | 1485 | 402 |
| Varanasi | 169516 | 132045 | 5083 | 1298 | 8311 | 5329 | 838 | 2219 | 6237 | 6611 | 765 | 780 |
| Visakha. | - | - | - | - | - | - | - | - | - | - | - | - |
| Total | 9315093 | 6527438 | 374186 | 75158 | 1505160 | 115296 | 90539 | 127120 | 352307 | 77192 | 29805 | 40892 |

Note: A: Included in cars.
        - : Not Available.

## Category/Surface-wise Road Length in India
(31st March 1999)

(In Kms.)

| India/ Category | Total | Surfaced | | | | Un- surfaced | | |
|---|---|---|---|---|---|---|---|---|
| | | Total | W.B.M. | B.T. | C.C. | Total | Motorable | Non-Motorable |
| **A. Highways** | **2017600** | **1217313** | **425373** | **790673** | **1267** | **800287** | **390418** | **409869** |
| a. PWD Roads | 989190 | 841537 | 188381 | 652867 | 289 | 147653 | 112844 | 34809 |
| (i) National Highways | 49585 | 49368 | 59 | 49297 | 12 | 217 | 180 | 37 |
| (ii) State Highways | 137950 | 135679 | 4071 | 131519 | 89 | 2271 | 1982 | 289 |
| (iii) Other PWD Roads | 801655 | 656490 | 184251 | 472051 | 188 | 145165 | 110682 | 34483 |
| b. Panchayat Raj Roads | 1028410 | 375776 | 23992 | 137806 | 978 | 652634 | 277574 | 375060 |
| (i) Zilla Parishad Roads | 456666 | 250995 | 170941 | 79945 | 109 | 205671 | 109531 | 96140 |
| (ii) Village Panchayat Roads | 425486 | 69485 | 29501 | 39965 | 19 | 356001 | 109391 | 246610 |
| (iii) CD/Panchayat Samiti Road | 146258 | 55296 | 36550 | 17896 | 850 | 90962 | 58652 | 32310 |
| **B. Urban Roads** | **237866** | **180558** | **49187** | **122816** | **8555** | **57308** | **48710** | **8598** |
| (i) Municipal Roads | 214475 | 159169 | 46596 | 104355 | 8218 | 55306 | 46952 | 8354 |
| (ii) MES Roads | 11883 | 11725 | 238 | 11463 | 24 | 158 | 144 | 14 |
| (iii) Railway Roads | 10282 | 8464 | 2309 | 5898 | 257 | 1818 | 1596 | 222 |
| (iv) Major Port Roads | 841 | 818 | 19 | 759 | 40 | 23 | 15 | 8 |
| (v) Minor Port Roads | 385 | 382 | 25 | 341 | 16 | 3 | 3 | - |
| **C. Project Roads** | **270523** | **50758** | **25753** | **24641** | **364** | **219765** | **137071** | **82694** |

| (i) Forest Deptt. | 162508 | 9666 | 6522 | 2971 | 173 | 152842 | 101044 | 51798 |
|---|---|---|---|---|---|---|---|---|
| (ii) Irrigation Deptt. | 74017 | 19051 | 11851 | 7143 | 57 | 54966 | 27530 | 27436 |
| (iii) Electricity Deptt. | 4657 | 4049 | 960 | 2982 | 107 | 608 | 509 | 99 |
| (iv) Sugar Cane Authority | 22972 | 12293 | 5309 | 6965 | 19 | 10679 | 7381 | 3298 |
| (v) Coal Mines Authority | 3923 | 3493 | 872 | 2613 | 8 | 430 | 367 | 63 |
| (vi) Steel Authority | 2446 | 2206 | 239 | 1967 | - | 240 | 240 | - |
| **India** | **2525989** | **1448629** | **500313** | **938130** | **10186** | **1077360** | **576199** | **501161** |

## State-wise Density of Rail Routes in India
(As on 31.3.2001)

| States/UTs | Route Kms. Per Lakh of Population | Route Kms. Per ' 000 Sq. Kms. |
|---|---|---|
| Delhi | 1.45 | 134.63 |
| Chandigarh | 0.86 | 67.89 |
| Punjab | 8.65 | 41.73 |
| West Bengal | 4.56 | 41.26 |
| Bihar | 4.15 | 36.55 |
| Uttar Pradesh | 5.16 | 35.93 |
| Haryana | 7.34 | 35.00 |
| Tamil Nadu | 6.74 | 32.21 |
| Assam | 9.45 | 32.08 |
| Gujarat | 10.50 | 27.10 |
| Kerala | 3.30 | 27.02 |
| Pondicherry | 1.14 | 22.56 |
| Jharkhand | 6.68 | 22.54 |
| Goa | 5.16 | 18.72 |
| Andhra Pradesh | 6.78 | 18.67 |
| Maharashtra | 5.64 | 17.74 |
| Rajasthan | 10.49 | 17.32 |
| Madhya Pradesh | 7.93 | 15.52 |
| Karnataka | 5.64 | 15.51 |
| Orissa | 6.29 | 14.83 |
| Chhattisgarh | 5.68 | 8.73 |
| Uttranchal | 4.20 | 6.37 |
| Himachal Pradesh | 4.42 | 4.83 |
| Tripura | 1.40 | 4.26 |
| Nagaland | 0.65 | 0.78 |
| Jammu & Kashmir | 0.95 | 0.43 |
| Mizoram | 0.17 | 0.07 |
| Manipur | 0.06 | 0.06 |
| Arunachal Pradesh | 0.12 | 0.02 |
| Meghalaya | 0 | 0 |
| Sikkim | 0 | 0 |
| Andaman & Nicobar Islands | 0 | 0 |
| Dadra & Nagar Haveli | 0 | 0 |
| Daman & Diu | 0 | 0 |
| Lakshadweep | 0 | 0 |
| **India** | **135.56** | **700.36** |

Based on real life data we have taken some sample token data in order to build some statistical analysis to make some indicative measure between heuristic algorithmic approaches against deterministic approach

| Cars | Motor Cycle | LCV | Buses | Trucks | Vehicles | Rickshaw | SUM |
|------|-------------|-----|-------|--------|----------|----------|-----|
| 840 | 895 | 268 | 209 | 2855 | 3014 | 551 | 8632 |
| 1265 | 815 | 493 | 707 | 1267 | 836 | 2363 | 7746 |
| 1004 | 464 | 188 | 583 | 373 | 118 | 2449 | 5179 |
| 4164 | 2536 | 667 | 203 | 1012 | 126 | 9334 | 18042 |
| 3830 | 1481 | 433 | 226 | 636 | 7 | 1541 | 8154 |
| 1757 | 3181 | 295 | 556 | 1715 | 598 | 1780 | 9882 |
| 870 | 1606 | 286 | 333 | 2413 | 821 | 1875 | 8204 |
| 4643 | 3470 | 1910 | 2122 | 1155 | 167 | 4637 | 18104 |
| 662 | 554 | 319 | 527 | 768 | 35 | 2775 | 5640 |
| 746 | 695 | 270 | 194 | 486 | 237 | 1066 | 3694 |

**Sample size: 20**

| Cars | Motor Cycle | LCV | Buses | Trucks | Vehicles | Rickshaw | Expectation |
|------|-------------|-----|-------|--------|----------|----------|-------------|
| 2 | 2 | 1 | 0 | 7 | 7 | 1 | 5.4 |
| 3 | 2 | 2 | 2 | 3 | 2 | 6 | 3.5 |
| 4 | 2 | 1 | 2 | 1 | 0 | 9 | 5.35 |
| 5 | 3 | 1 | 0 | 1 | 0 | 10 | 6.8 |
| 9 | 4 | 1 | 1 | 2 | 0 | 4 | 5.95 |
| 4 | 6 | 1 | 1 | 3 | 1 | 4 | 4 |
| 2 | 4 | 1 | 1 | 6 | 2 | 5 | 4.35 |
| 5 | 4 | 2 | 2 | 1 | 0 | 5 | 3.75 |
| 2 | 2 | 1 | 2 | 3 | 0 | 10 | 6.1 |
| 4 | 4 | 1 | 1 | 3 | 1 | 6 | 4 |

**Sample size: 25**

| Cars | Motor | LCV | Buses | Trucks | Vehicles | Rickshaw | Expectation |
|------|-------|-----|-------|--------|----------|----------|-------------|
| 2 | 3 | 1 | 1 | 8 | 8 | 2 | 5.88 |
| 4 | 3 | 2 | 2 | 4 | 2 | 8 | 4.68 |
| 5 | 2 | 1 | 3 | 2 | 0 | 12 | 7.48 |
| 6 | 4 | 1 | 0 | 1 | 0 | 13 | 8.92 |
| 12 | 5 | 1 | 1 | 1 | 0 | 5 | 7.88 |
| 4 | 8 | 1 | 1 | 4 | 2 | 5 | 5.08 |
| 3 | 5 | 1 | 1 | 7 | 2 | 6 | 5 |
| 6 | 5 | 3 | 3 | 2 | 0 | 6 | 4.76 |
| 3 | 2 | 1 | 2 | 3 | 2 | 12 | 7 |
| 5 | 5 | 2 | 1 | 3 | 2 | 7 | 4.68 |

**Sample size: 30**

| Cars | Motor Cycle | LCV | Buses | Trucks | Vehicles | Rickshaw | Expectation |
|---|---|---|---|---|---|---|---|
| 3 | 3 | 1 | 1 | 10 | 10 | 2 | 7.47 |
| 5 | 3 | 2 | 3 | 5 | 3 | 9 | 5.4 |
| 6 | 3 | 1 | 3 | 2 | 1 | 14 | 8.53 |
| 7 | 4 | 1 | 0 | 2 | 0 | 16 | 10.87 |
| 14 | 5 | 2 | 1 | 2 | 0 | 6 | 8.87 |
| 5 | 10 | 1 | 2 | 5 | 2 | 5 | 6.13 |
| 3 | 6 | 1 | 1 | 9 | 3 | 7 | 6.2 |
| 8 | 6 | 3 | 4 | 1 | 0 | 8 | 6.33 |
| 4 | 3 | 2 | 3 | 3 | 0 | 15 | 9.07 |
| 6 | 6 | 2 | 2 | 3 | 2 | 9 | 5.8 |

**Sample size: 40**

| Cars | Motor | LCV | Buses | Trucks | Vehicles | Rickshaw | Expectation |
|---|---|---|---|---|---|---|---|
| 4 | 4 | 1 | 1 | 13 | 14 | 3 | 10.2 |
| 7 | 4 | 3 | 4 | 6 | 4 | 12 | 7.15 |
| 8 | 4 | 1 | 5 | 2 | 1 | 19 | 11.8 |
| 9 | 6 | 1 | 0 | 2 | 1 | 21 | 14.1 |
| 19 | 7 | 2 | 1 | 3 | 0 | 8 | 12.2 |
| 7 | 13 | 1 | 2 | 7 | 3 | 7 | 8.25 |
| 4 | 8 | 1 | 2 | 12 | 4 | 9 | 8.15 |
| 10 | 8 | 4 | 5 | 3 | 0 | 10 | 7.85 |
| 5 | 4 | 2 | 4 | 5 | 0 | 20 | 12.15 |
| 8 | 8 | 3 | 2 | 4 | 3 | 12 | 7.75 |

**Sample size: 50**

| LCV | Buses | Trucks | Vehicles | Rickshaw | Expectation |
|---|---|---|---|---|---|
| 2 | 1 | 17 | 17 | 3 | 12.84 |
| 3 | 5 | 8 | 6 | 15 | 8.96 |
| 2 | 6 | 3 | 1 | 24 | 14.84 |
| 2 | 1 | 2 | 0 | 26 | 17.56 |
| 3 | 1 | 4 | 1 | 9 | 14.36 |
| 1 | 3 | 9 | 3 | 9 | 10.36 |
| 2 | 2 | 15 | 5 | 11 | 10.08 |
| 5 | 6 | 3 | 0 | 13 | 10.16 |
| 3 | 5 | 6 | 0 | 25 | 15.12 |
| 4 | 3 | 7 | 3 | 14 | 9.2 |

| Sample Size | Standard Deviation |
|---|---|
| 20 | 15.47 |
| 20 | 16.23 |
| 20 | 17 |
| 20 | 17.96 |
| 20 | 18.94 |
| 20 | 19.72 |
| 20 | 20.71 |
| 20 | 21.69 |
| 20 | 22.49 |
| 20 | 23.28 |

| Sample Size | Standard Deviation |
|---|---|
| 25 | 13.84 |
| 25 | 14.52 |
| 25 | 15.2 |
| 25 | 16.07 |
| 25 | 16.94 |
| 25 | 17.64 |
| 25 | 18.52 |
| 25 | 19.4 |
| 25 | 20.11 |
| 25 | 20.83 |

| Sample Size | Standard Deviation |
|---|---|
| 30 | 12.63 |
| 30 | 13.25 |
| 30 | 13.88 |
| 30 | 14.67 |
| 30 | 15.46 |
| 30 | 16.1 |
| 30 | 16.91 |
| 30 | 17.71 |
| 30 | 18.36 |
| 30 | 19.01 |

| Sample Size | Standard Deviation |
|:---:|:---:|
| 40 | 10.94 |
| 40 | 11.48 |
| 40 | 12.02 |
| 40 | 12.7 |
| 40 | 13.39 |
| 40 | 13.95 |
| 40 | 14.64 |
| 40 | 15.34 |
| 40 | 15.9 |
| 40 | 16.46 |

| Sample Size | Standard Deviation |
|:---:|:---:|
| 50 | 9.78 |
| 50 | 10.26 |
| 50 | 10.75 |
| 50 | 11.36 |
| 50 | 11.98 |
| 50 | 12.47 |
| 50 | 13.1 |
| 50 | 13.72 |
| 50 | 14.22 |
| 50 | 14.73 |

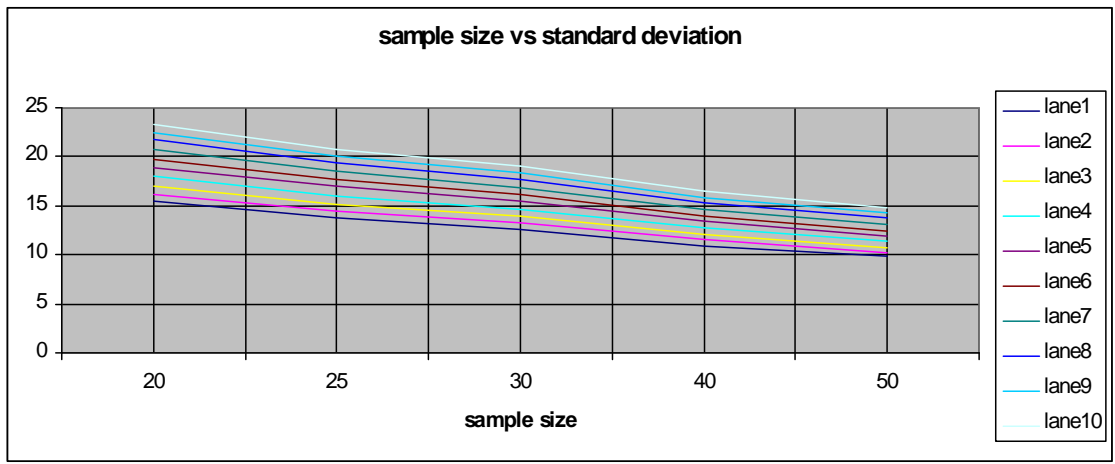

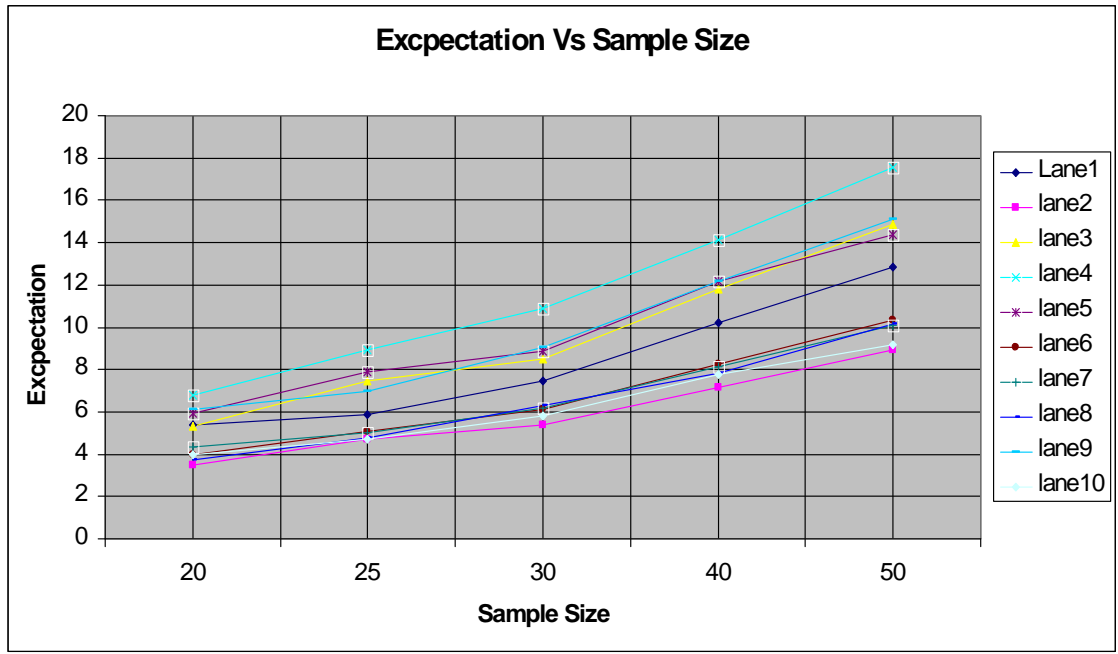

## 6 CONCLUSIONS AND FUTURE SCOPE

The article presented through this paper mainly emphasize on optimal usage of lanes using threshold information as knowledge base, but at the cost of transitions, because in real life scenario transitions may be too high, hence our future effort will be certainly in this direction.
In this article amount of time taken to transit between lanes has been considered cannot be ignored. The cumulative sum of transition time between lanes in real world problem contributed much in optimality of the proposed solution.

Bio inspired algorithms (like swarm intelligence) has been used with population information as knowledge base in our previous works, but partial modification of the stated concept taking threshold level information of the respective lanes will certainly be taken into consideration but implementation and formulation of algorithms along with optimality in transition, there by optimizing various aspects of traffic movement in real world will be considered in our future effort.

# Authors

**Dr. Prasun Ghosal** is currently associated with Bengal Engineering and Science University as an Assistant Professor. He has completed PhD (2011) from Bengal Engineering and Science University, M.Tech. (2005) as well as B.Tech. (2002) in Radio Physics and Electronics Engineering from Institute of Radio Physics and Electronics, University of Calcutta, India. He is also an Honours Graduate (major in Physics) under University of Calcutta. He has received Young Scientist Research Award for the year 2010-11 from Indian Science Congress Association. He is also recipient of several Best paper awards. His research interests include VLSI Physical Design: Algorithms and Applications, Network on Chip: Architectures and Algorithms, Embedded Systems: Architectures and Applications, Quantum Computing, Circuits, and Cryptography. He has contributed around 10 research articles in several peer reviewed international journals and around 40 in peer reviewed international conferences. Besides a copyright application he has also contributed towards several book chapters. He has carried out several sponsored research projects funded by AICTE, DIT, MCIT, Govt. of India, IEI etc.

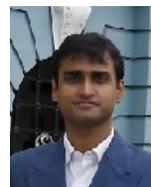

**Arijit Chakraborty** is currently working in Heritage Institute of Technology. He is a post graduate (M.sc) in IT and also done his M. Tech (IT) from Bengal Engineering and Science University, Shibpur, India. His research interests include Soft Computing, Bio Inspired Algorithms.

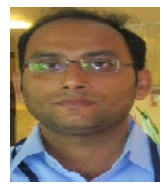

**Sabyasachee Banerjee,** received B. Tech. degree in computer science from Government College of Engineering and Textile Technology, Shrirampore, under WBUT, WB, India in 2008 and M. E. in Computer Science from Bengal Engineering and Science University, Shibpur, WB, India in 2011. He is currently with the Department of Computer Science and Engineering, Heritage Institute of Technology, Kolkata, WB, India. His current research interests include Soft Computing, VLSI Physical Design algorithms and optimization.

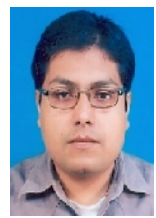

Satabdi Barman is currently working in Heritage Institute of Technology for last 5 years. She has done her B.tech and M.tech from Vidyasagar University and West Bengal University of Technology respectively. Her research area Includes Soft Computing, Bio Inspired Algorithms, Real Time System.

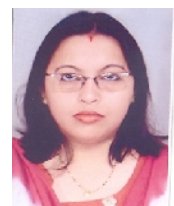